\documentstyle[prb,aps,twocolumn,psfig]{revtex}

\begin{document}
\wideabs{
\title{Kinetic energy of solid neon by Monte Carlo\\
       with improved Trotter- and finite-size extrapolation}
\author{Alessandro Cuccoli\cite{email}, Alessandro Macchi\cite{email},
        Gaia Pedrolli\cite{email}, and Valerio Tognetti\cite{email}}
\address {Dipartimento di Fisica dell'Universit\`a di Firenze
          and Istituto Nazionale di Fisica della Materia (INFM),\\
          Largo E. Fermi~2, I-50125 Firenze, Italy}
\author{Ruggero Vaia\cite{email}}
\address { Istituto di Elettronica Quantistica
           del Consiglio Nazionale delle Ricerche,\\
           via Panciatichi~56/30, I-50127 Firenze, Italy}
\date{Phys. Rev. B {\bf 56}, 51 (1997)}
\maketitle
\begin{abstract}
The kinetic energy of solid neon is calculated by a path-integral Monte
Carlo approach with a refined Trotter- and finite-size extrapolation. These
accurate data present significant quantum effects up to temperature
$T=20$~K. They confirm previous simulations  and are consistent with recent
experiments.
\end{abstract}
}
In a previous work \cite{kk} we reported theoretical results about the
average kinetic energy of rare gas solids (krypton, argon and neon),
modeled by a Lennard-Jones (LJ) interaction. For heavier crystals the
thermodynamics was approached by means of the effective-potential
method \cite{G-T,F-K}. This approach allows us the use of all
classical methods through the construction of an approximate effective
classical phase-space distribution (see for details \cite{rev95}).
Monte Carlo  simulations \cite{Horton1,Mac1,Wallis,Acocella,kk} with the
effective potential were favourably compared with path-integral Monte Carlo
(PIMC) simulations, and also applied to argon \cite{Mac1}, reproducing very
well the experimental density and specific heat \cite{exp-argon}.

In spite of their large mass, krypton and argon show relevant quantum
effects at easily accessible temperatures; for instance, the average
kinetic energy is much larger than its corresponding classical value. Our
calculations of the average kinetic energy of argon suggested the
realization of neutron Compton scattering (NCS) experiments, whose
outcomes have been later found in perfect agreement with our predictions
\cite{Simmons}.
For increasing value of the quantum coupling, anharmonic second order
corrections arising from the odd part of the potential were found relevant
\cite{Wallis} and have been recently inserted in the effective potential
formalism \cite{prlhorton}.

For neon, where the strong anharmonicity shows up even in the ground state
\cite{Muser}, we preferred to resort to path-integral Monte Carlo (PIMC)
simulations, with a refined Trotter extrapolation \cite{kk}; the procedure
consists in adding to the PIMC data the contribution of the harmonic
approximation after the subtraction of the corresponding finite Trotter
data.

The available experimental data for the kinetic energy of solid neon
obtained by NCS experiments \cite{Peek}, were in evident disagreement with
our PIMC results; all theoretical data were lower than the experimental
ones. The reason was not clear at all. Since the experiments with argon
\cite{Simmons} were not yet performed, we supposed that the LJ
potential were an over-simplification of the true potential and that
many-body interactions could play an important role.

In a recent work Timms {\em et al.} \cite{Timms} report new NCS
measurements of the kinetic energy of solid neon, which differ from the
previous ones \cite{Peek}. Both experiments were carried out and analyzed
in the regime of the so-called {\em impulse approximation}, which becomes
exact when the energy and the momentum transferred to the sample are
infinite. Timms {\em et al.} reached higher momentum and energy transfers,
and they claim that this improvement has made the final-state-effect
corrections to the observed longitudinal Compton profile irrelevant: hence,
the data analysis had less sources of uncertainty and was thus more
reliable. They also did new PIMC simulations, using different potentials
(Aziz and LJ) in order to definitively establish whether the
original disagreement between theory and experiment was due to a too rough
model potential: it turned out that the computed kinetic energy depends
very weakly on the model potential. Their experimental and PIMC data are
consistent between themselves, and also confirm the validity of our previous
simulations.

In a recent paper \cite{trick} we have developed a systematic method for
improving the Trotter number extrapolation of PIMC data; in addition, we
have recently extended this procedure in order to take into account also
the effect of the finite size of the simulation box \cite{trick2}, so that
we are now able to obtain much more accurate results.

The PIMC method is based upon the semi-group property of the
density matrix
\begin{equation}
 \rho({\mbox{\boldmath$q$}}^\prime,{\mbox{\boldmath$q$}};\beta)
 = \int d{\mbox{\boldmath$q$}}_{P-1}...d{\mbox{\boldmath$q$}}_{1}
 \rho({\mbox{\boldmath$q$}}^\prime,{\mbox{\boldmath$q$}}_{P-1};\tau)...
 \rho({\mbox{\boldmath$q$}}_1,{\mbox{\boldmath$q$}};\tau),
\label{semigroup}
\end{equation}
where $\beta=1/T$ and $\tau=\beta/P$, P being the Trotter number. In order
to make the above formula of practical use, the density matrix element
$\rho({\mbox{\boldmath$q$}}_\ell,{\mbox{\boldmath$q$}}_{\ell-1};\tau)$ is
usually taken in the lowest high-temperature approximation, giving rise to
the so-called {\em primitive} action. The latter tends to the exact density
matrix as $P\to\infty$. This is the formalism we are dealing with here. In
order to get values of the averages of physical observables, many
simulations at different values of $P$ must be performed, and then the data
must be extrapolated as $P\to\infty$. The more ``quantum'' is the system,
the larger must be the maximum value of $P$. The finite $P$ estimates
$G(P)$ of the averages can be expanded as
$G(P)=G(\infty)+g_2/P^2+g_4/P^4+\cdots$ \cite{Suzuki}, and
frequently the term in $1/P^2$ is not sufficient for a satisfactory fit.

The method we suggested \cite{trick} overcomes this problem. The idea is to
take advantage of the fact that the thermodynamics of a harmonic system can
be obtained analytically, even at finite $P$ (see
Ref.~\onlinecite{Wolynes}); nevertheless,
such a system shows a very strong dependence on $P$, and the results
obtained at finite $P$ will not be close to those at $P=\infty$ unless the
condition $P\gg{f}=\beta\hbar\omega/2$ is fulfilled for any system's
frequency $\omega$. However, at low temperatures, when quantum effects are
most important (in the temperature region where one should use the highest
values of $P$), the harmonic approximation (HA) of a solid system is surely
meaningful, though rough. Indeed, as $T\to{0}$, harmonic excitations play a
very important role in the thermodynamics. As long as the ``quantum
character'' of the system increases, this becomes less and less true, but
the self-consistent HA (SCHA) \cite{Koehler,Werthamer} eventually allows to
recover a simple harmonic-like system whose behaviour is very similar to
that of the real system.

Our idea is to improve the extrapolation as $P\to\infty$ in PIMC
simulations accounting for the $P$ dependence of the harmonic contributions
to the PIMC estimates of physical observables. The procedure consists in
adding to the rough PIMC data $G(P)$ the deviation from the $P=\infty$
estimates calculated for the SCHA of the system:
\begin{equation}
 G_{SC}(P) = G(P) + \left[ G_{SC}^{(h)}(\infty) - G_{SC}^{(h)}(P) \right]~.
\label{P_trick}
\end{equation}
In such a way, the improved estimates $G_{SC}(P)$ will show a much 
weaker dependence on the Trotter number $P$, the scaling behaviour
in $1/P^2$ is reached earlier and the maximum Trotter number necessary
to get the correct asymptotic result is lower. We remember that as $P$
increases the computer time grows both because of the larger simulation box
and of the worse statistics.

Another important point which has not yet been deeply investigated in
relation to quantum simulations is the dependence of the data on the
simulation box size. It is well known that for systems undergoing a phase
transition the finite size of the simulated sample has dramatic effects,
because in the critical region the correlation length diverges, and in
order to simulate such a system particular procedures known as
``finite-size scaling'' must be used. In the classical case, if the system
is far from a phase transition, the problem of how to reach the
thermodynamic limit regime is in general easily overcome, without using
enormous samples (size effects, with modern computer capability, are in
general not a problem). However, dealing with quantum systems, subtle
phenomena can occur: for instance,  in $d$-dimensional lattices the
discreteness of the Brillouin zone due  to the finite particle number
introduces a nonphysical gap into the  dispersion curve which gives rise,
at low temperature, to an exponential  behaviour of the specific heat,
instead of the correct $T^d$ scaling Bloch law, obtained for a linear
dispersion of the soft modes. This effect can be observed in simulations
\cite{Muser}.

Following the idea of Eq. (\ref{P_trick}) we suggest \cite{trick2} to
correct the raw PIMC data, at the SCHA level, also with respect to their
dependence on $N$, the number of particles composing the actually simulated
sample,
\begin{equation}
 G_{SC}(P,N) = G(P,N) +
 \left[ G_{SC}^{(h)}(\infty,\infty) - G_{SC}^{(h)}(P,N) \right]~.
\label{N_trick}
\end{equation}
Preliminary tests \cite{trick2} made on 1-d nonlinear systems confirm
that even with a chain composed by very few particles it is possible to get
the thermodynamic limit of the averages of observables only adding
to the raw simulation data this harmonic-like correction, practically
making the extrapolation as $N\to\infty$ unnecessary.

To model solid neon, we considered a fcc lattice composed by $N$ particles
(labeled with 3-d indices {\bf i, j}) interacting through a pairwise
potential,
\begin{equation}
 V({\mbox{\boldmath$q$}}) = \frac{1}{2} \sum_{\bf ij}
 v\big(|{\mbox{\boldmath$q$}}_{{\bf i}}-{\mbox{\boldmath$q$}}_{\bf j}|\big)~.
\label{two-body}
\end{equation}
Several choices, as we know, are possible for the model potential. Since {\em
Timms et al.} \cite{Timms} showed that the dependence  of the kinetic
energy on the potential is not critical, we chose to model  our system by
the LJ 12-6 potential
$v(r)=4\varepsilon\big[(\sigma/r)^{12}-(\sigma/r)^{6}\big]$ with the
potential parameters $\varepsilon$ and $\sigma$ taken from the 
literature \cite{neon_LJ_par} ($\varepsilon=36.68$~K and
$\sigma=2.787$~\AA). We neglect the dynamic effect of the interactions
beyond nearest-neighbours, whose contribution to the potential energy
is taken into account by a static-lattice approximation.
We performed constant-density simulations evaluating the pressure within
each run. The density was adjusted in such a way to get a practically
vanishing  pressure, (the pressure is always less than 
$0.07\,p^*\approx{15}$~atm, being $p^*=\varepsilon/\sigma^3$ the 
characteristic pressure) in order to best reproduce the experimental
settings; the zero-pressure densities turned out to be very close to the
experimental ones. The sample was an fcc lattice of $108$ atoms with
periodic boundary conditions;
in order to test the above described finite-size correction scheme
we made test runs changing the box size up to $864$ particles. We used the
Metropolis algorithm, with both single- and many-particle moves. The
maximum Trotter number $P$ was $48$. Each run consisted of $200,000$ steps
per particle for equilibration followed by $1,200,000$ further steps,
during which the averages were accumulated every $5$ steps. For each run we
estimated the statistical uncertainty, taking into account the MC
correlation times; these vary with $P$ and $N$ and never exceed 
400 steps.

In order to make finite-Trotter and finite-size harmonic corrections in the
spirit described above, we need the SCHA potential,
\begin{equation}
 V_0({\mbox{\boldmath$q$}}) = \frac{1}{2} N z w + \frac{1}{2} m
 \sum_{{\bf i}\alpha} \sum_{{\bf j}\beta}
 {\Omega^2}_{\bf i j}^{\alpha \beta}
 \xi_{\bf i}^{\alpha} \xi_{\bf j}^{\beta}~,
\label{V_0(q)}
\end{equation}
where $\xi_{\bf i}^{\alpha}=q_{\bf{i}}^{\alpha}-q_{0,{\bf i}}^{\alpha}$ and
${\bf{q}}_{0,{\bf{i}}}$ is the equilibrium position of the $i-th$ particle,
which is fixed, being determined by the particle density. $z=12$ is the
coordination number and $w$ and ${\Omega^2}_{\bf{ij}}^{\alpha\beta}$ are
adjustable parameters determined imposing that the average of the actual
potential $V({\mbox{\boldmath$q$}})$, and of its first and second
derivatives are equal to the corresponding averages obtained for
$V_0({\mbox{\boldmath$q$}})$: all averages are performed using the 
finite-$P$ density distribution corresponding to
$V_0({\mbox{\boldmath$q$}})$, which is a gaussian. A shorthand way of
expressing these gaussian averages as a formal power series turns out to be
very useful in this case, namely
$\big\langle{f}(\mbox{\boldmath$\xi$}_{\bf{id}})\big\rangle_0\equiv
\exp({\cal{D}}^{\alpha\beta}\partial_{\alpha}\partial_{\beta}/2)\,f({\bf{0}})$
(summation over repeated indices) where
$\mbox{\boldmath$\xi$}_{\bf{id}}\equiv
\mbox{\boldmath$\xi$}_{{\bf{i}}+{\bf{d}}}{-}
\mbox{\boldmath$\xi$}_{\bf{i}}$ ({\bf{d}} labels the nearest-neighbour
displacements) and
${\cal D}^{\alpha\beta}=\big\langle
\xi_{\bf{id}}^{\alpha}\xi_{\bf{id}}^{\beta}\big\rangle_0$
is the variance matrix of the gaussian distribution.

Since $v(r)\to\infty$ as $r$ approaches $0$, averages like
$\big\langle{v}\big(|{\bf{d}}{+}\mbox{\boldmath$\xi$}_{\bf{id}}|
\big)\big\rangle_0$
would diverge. This is an artifact of the harmonic approximation: the
gaussian is small but nonzero at $r=0$, where the true distribution would
vanish; the formal power series
$\exp({\cal D}^{\alpha\beta}\partial_{\alpha}\partial_{\beta}/2)
V({\mbox{\boldmath$q$}})$
is then only asymptotic. However, the nonphysical contributions from the
potential core can be simply eliminated by truncating the series.
Hence we can expand the averaged potential and its derivatives up to second
order in the ${\cal D}'s$, finally obtaining the following SCHA equations
\begin{eqnarray}
 {m\,{\Omega^2}_{\bf k}^{\nu} \over2}&=&
 \tilde v''\big(\nu_{\bf{k}}^2+A_{\bf{k}}^\nu\big)+
 {\tilde v'\over d}\big(\nu_{\bf{k}}^2-A_{\bf{k}}^\nu\big)
\\
 & & \hspace{-10truemm} w = v +
 {1\over4}\bigg[ \big(v''+\tilde v''\big){\cal{D}}_\|+
 \Big({v'\over d}+{\tilde v'\over d}\Big) {\cal{D}}_\perp\bigg]
\nonumber \\
 & &\hspace{30truemm}
 -\frac{1}{Nz}\sum_{{\bf k}\nu}m{\Omega^2}_{\bf k}^{\nu}\alpha_{\bf k}^{\nu}~,
\end{eqnarray}
where $v\equiv{v}(d)$, $v'\equiv{v'}(d)$ and so on; $d\equiv|{\bf d}|$ is
the nearest-neighbour distance. The indices ${\bf{k}}$ (wavevector) and
$\nu$ (polarization) label the normal modes, obtained by diagonalizing
${\Omega^2}_{\bf ij}^{\alpha\beta}$ to the eigenfrequencies
${\Omega^2}_{\bf{k}}^\nu$. In particular,
$4\nu_{\bf{k}}^2=\sum_{\bf{d}}[1-\cos({\bf{k}}\cdot{\bf{d}})]$ and
$A_{\bf{k}}^\nu$ results from the polarization diagonalization.
The quantity \cite{trick,Wolynes}
\begin{equation}
 \alpha_{\bf{k}}^\nu
 =\big\langle \xi_{\bf{k}}^\nu \xi_{\bf{k}}^\nu \big\rangle_0
 ={\hbar\over 2m\Omega_{\bf{k}}^\nu}
 {\coth(P\mu_{\bf{k}}^\nu)\over\cosh\mu_{\bf{k}}^\nu}
\end{equation}
is the normal coordinate mean square fluctuation at finite $P$, with
$\sinh(\mu_{\bf{k}}^\nu)\equiv\beta\hbar\Omega_{\bf{k}}^\nu/(2P)$~;
the corresponding limit for $P\to\infty$ is easily recovered. Moreover,
\begin{eqnarray}
 \tilde v'' &=& v''+ {v^{(4)}\over2}{\cal{D}}_\| +\left(
 {v'''\over 2d}-{v''\over d^2} +{v'\over d^3}\right){\cal{D}}_\perp
\\
 \tilde v' &=& v'+
 \left({v'''\over2} - {v''\over d} + {v'\over d^2}\right){\cal{D}}_\|
 +\left({v''\over d} - {v'\over d^2} \right) {\cal{D}}_\perp \, .
\end{eqnarray}
${\cal{D}}_\|$ and ${\cal{D}}_\perp$ are the mean square fluctuations of
the components of ${\mbox{\boldmath$\xi$}}_{\bf{id}}$, parallel and
orthogonal to ${\bf{d}}$, respectively. Due to the symmetry properties of
the fcc lattice, the $108$ ${\cal D}$'s reduce indeed to three only, and
making a further isotropy approximation we assume that the two transverse
components are equal, $ {\cal D}_{\perp,1} \simeq {\cal D}_{\perp,2} \simeq
{\cal D}_{\perp} $.
\begin{equation}
 {\cal D}_{\|,\perp}={1\over3N}\sum_{{\bf{k}}\nu}
 \big(\nu_{\bf{k}}^2\pm A_{\bf{k}}^\nu\big)\,\alpha_{\bf{k}}^\nu~.
\end{equation}
The SCHA finite$-P$ estimates can be obtained as logarithmic
derivatives of the partition function
\begin{equation}
{\cal Z}_0(P,N) = e^{-\beta Nzw/2}~
\prod_{ {\bf k}\nu} \left[ \, 2 \, \sinh(P \mu_{{\bf k}}^\nu) \right]^{-1}~.
\label{Z-finite_P}
\end{equation}
The well-known partition function for a system of quantum harmonic
oscillators is recovered as $P \to\infty$. In order to get the finite-$N$
values we use the discrete mesh in the Brillouin zone corresponding to that
value of $N$; the thermodynamic limit is obtained by the special points
method. The SCHA kinetic energy is
${\cal K} = (2N)^{-1}\sum_{{\bf k}\nu}\,
m {\Omega^2}_{\bf k}^{\nu}\alpha_{\bf k}^{\nu}$
where $\alpha_{\bf k}^{\nu}=\alpha_{\bf{k}}^\nu(P,N)$.
In this way, we are able to get both finite $P$ and finite $N$
corrections.

\begin{figure}[hbt]
\centerline{\psfig{bbllx=15mm,bblly=155mm,bburx=188mm,bbury=285mm,%
figure=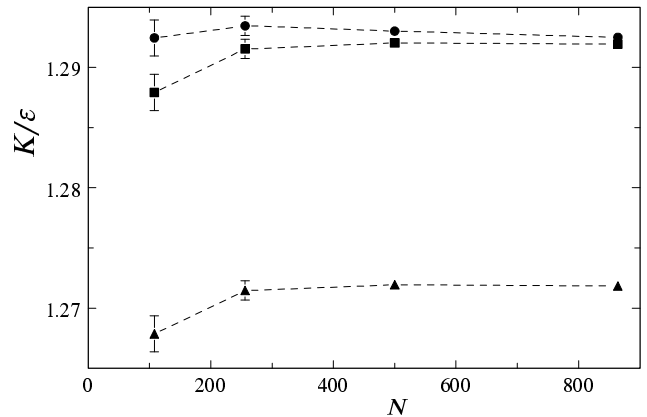,width=86mm,angle=0}}
\caption{
Reduced kinetic energy ${\cal K}/\epsilon$ vs. particle number $N$.
The Trotter number is $P = 8$, the temperature is $T=20$~K and the reduced
density is $\rho/\rho^* = 0.945 $.
$\rho^*\equiv{m}/\sigma^3=1.5479$~g\,cm$^{-3}$ is the characteristic
density. The triangles are raw PIMC data, the squares are PIMC data plus
finite-Trotter corrections (\protect\ref{P_trick}) and the circles are PIMC
data plus finite-Trotter and  finite-size corrections
(\protect\ref{N_trick}). Error bars, when not shown, lie inside the
symbols. Lines are guides for the eye.
\label{figure1}
}
\end{figure}

\begin{figure}[hbt]
\centerline{\psfig{bbllx=15mm,bblly=155mm,bburx=188mm,bbury=285mm,%
figure=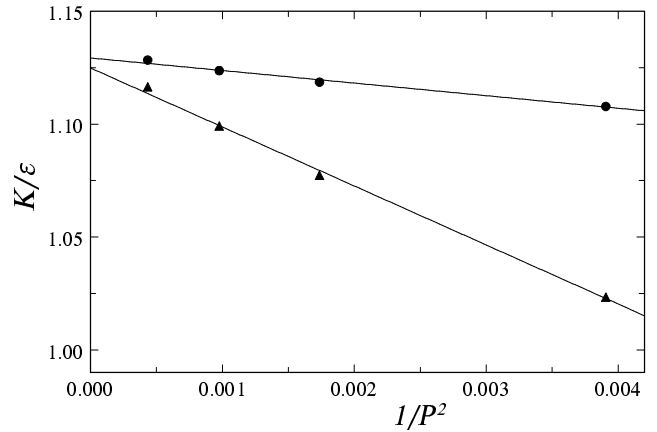,width=86mm,angle=0}}
\caption{
Reduced kinetic energy ${\cal K}/\epsilon$ vs. the inverse square of the
Trotter number $P$. The simulation box contains $108$ particles, the
temperature is $T=5$~K and the reduced density is $\rho/\rho^* = 0.968$.
The triangles are raw PIMC data, and the circles are PIMC data plus finite
Trotter and finite size corrections (\protect\ref{N_trick}). Error bars,
when not shown, lie inside the symbols. The lines are fits of
corresponding data.
\label{figure2}
}
\end{figure}

Through this analysis we have concluded that for the kinetic energy of neon
a simulation box with $108$ atoms and periodic boundary conditions is
barely large enough to mimic the thermodynamic limit behaviour ({\em i.e.}
the finite size corrections are of the order of the statistical error), as
it can be seen in Fig.~\ref{figure1} where it's shown how the finite-size 
corrections scheme works. The relative effect of finite-size and
finite-Trotter corrections is well seen also in Fig.~\ref{figure2}, where
it is shown how the finite-Trotter corrections work. The small difference
between the extrapolations at $P=\infty$ represents the size effect.

\begin{figure}[hbt]
\centerline{\psfig{bbllx=15mm,bblly=155mm,bburx=188mm,bbury=285mm,%
figure=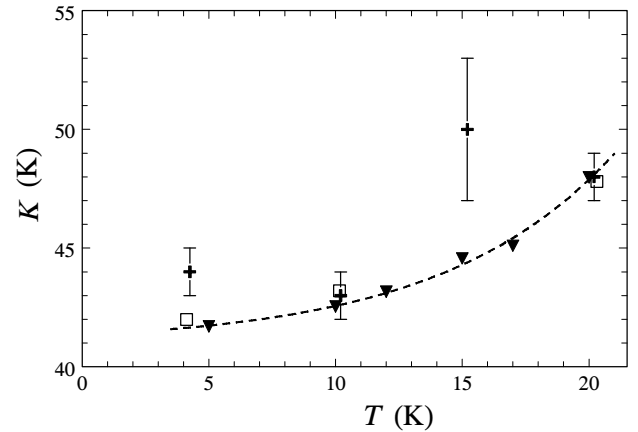,width=86mm,angle=0}}
\caption{
Kinetic energy ${\cal K}$ of solid neon vs. temperature $T$.
The crosses are experimental data from \protect\cite{Timms}, the open
squares are PIMC simulations data from \protect\cite{Timms} and the
triangles are our PIMC data with refined Trotter extrapolation. Error bars,
when not shown, lie inside the symbols. The line is a guide for the eye.
\label{figure3}
}
\end{figure}

The results of our simulations are shown in Fig.~\ref{figure3}: they are
consistent with the experimental data \cite{kk} and confirm the validity of
other PIMC simulations \cite{kk,Timms}.

In conclusion, the SCHA correction scheme gives an estimate on how large
both the finite-size and the finite-Trotter effects are. It is indeed
important to control how much the data are affected by the finiteness of
the simulation box, especially for observables which possibly show a
stronger $N$-dependence than the kinetic energy, such as the specific heat
or the frequency moments, or for systems with larger quantum coupling.
\medskip

We gratefully acknowledge useful discussions with P. Nielaba.

\end{document}